\documentclass{Interspeech2024}

\interspeechcameraready

\usepackage{hyperref}
\usepackage{multirow}
\usepackage{multicol}
\usepackage{comment}
\usepackage{subfig}
\usepackage{amsmath}
\usepackage{amssymb}
\usepackage[vlined]{algorithm2e}
\usepackage{url}
\usepackage{xcolor}
\usepackage{float}
\usepackage{algpseudocode}
\usepackage{graphicx}
\usepackage{booktabs}
\usepackage{subfig}

\usepackage{subcaption}
\usepackage{soul}
\usepackage[table]{colortbl}

\title{CoLLAB: A Collaborative Approach for Multilingual Abuse Detection}
\name[affiliation={1}]{Orchid}{Chetia Phukan*}
\name[affiliation={1}]{Yashasvi}{Chaurasia*}
\name[affiliation={1}]{Arun Balaji}{Buduru}
\name[affiliation={1,2}]{Rajesh}{Sharma}

\address{
  $^1$IIIT-Delhi, India\\
  $^2$University of Tartu, Estonia\\
  *equal contribution}
\email{orchidp@iiitd.ac.in}


\begin{document}

\maketitle

\begin{abstract}
In this study, we investigate representations from paralingual Pre-Trained model (PTM) for Audio Abuse Detection (AAD), which has not been explored for AAD. Our results demonstrate their superiority compared to other PTM representations on the ADIMA benchmark. Furthermore, combining PTM representations enhances AAD performance. Despite these improvements, challenges with cross-lingual generalizability still remain, and certain languages require training in the same language. This demands individual models for different languages, leading to scalability,  maintenance, and resource allocation issues and hindering the practical deployment of AAD systems in linguistically diverse real-world environments. To address this, we introduce \textbf{CoLLAB}, a novel framework that doesn't require training and allows seamless merging of models trained in different languages through weight-averaging. This results in a unified model with competitive AAD performance across multiple languages. \par  
\end{abstract}

\section{Introduction}

In today's digital era, online social media platforms, gaming communities, and digital environments have become indispensable aspects of modern life. Within these realms, audio plays a crucial role in facilitating real-time communication and fostering connections among users. However, alongside the advantages of audio communication in these settings, there is a growing concern regarding audio-abusive content (AAC). This refers to the malicious or inappropriate use of audio, encompassing behaviors such as verbal harassment, hate speech, and the dissemination of harmful or offensive material. The prevalence of AAC in digital spaces presents substantial challenges for both users and platform administrators. \par

Due to the challenges induced by AAC to online safety and well-being, audio-based abuse detection (AAD) has caught recent attention.  While abuse detection in other modalities like text \cite{davidson2017automated, madhu2023detecting} and visual \cite{alcantara2020offensive, gao2020offensive} content has garnered significant focus and development, AAD has not received comparable attention, despite its critical importance in safeguarding digital spaces. This gap underscores the urgent need for research and development efforts aimed at advancing AAD technologies to effectively mitigate the harmful impacts of audio-based abuse in online environments. \par

There has been growing interest in building effective AAD systems, for example, Gupta et al. \cite{9746718} explored representations from VGG trained on AudioSet and multilingual wav2vec2 models with Fully Connected Network (FCN), GRU, and LSTM as classifiers. Further, Sharon et al. \cite{sharon22_interspeech} showed that AAD performance can be enhanced by fusing audio and ASR transcribed textual representations. On the other hand, Thakran et al. \cite{thakran23_interspeech} showed that choosing the right acoustic cues by exploring multilingual and emotion recognition PTMs can lead to state-of-the-art (SOTA) performance in AAD without the need of textual representations also reverberated by Spiesberger et al. \cite{spiesberger23_interspeech}. 

AAD has benefitted significantly from the availability of different PTM representations, particularly those adept at capturing paralinguistic cues effectively. However, representations from paralingual PTM \cite{shor2022universal} which has shown SOTA performance in related tasks to AAD such as speech emotion recognition (SER), speaker identification (SI), and so on, haven't been explored for AAD. In this work, we investigate representations from paralingual PTM for AAD and the first study to do so, according to the best of our knowledge. We hypothesize that these representations will capture verbal cues such as pitch, tone, and intensity much more effectively in comparison to representations of other PTMs for AAD. To validate our hypothesis, we present a comprehensive comparative study of five PTM (TRILLsson,Whisper, MMS, WavLM, x-vector) representations including paralingual, multilingual, monolingual as well as speaker recognition PTM which are SOTA in respective tasks and highly potential candidates for improved AAD. \par

We also conducted an investigation by combining PTM representations to explore their potential for complementary behavior for more improved AAD, akin to observations in other tasks such as speech recognition \cite{arunkumar22b_interspeech}. We further sought to address the challenge of AAC being present in diverse languages as real-world digital environments are linguistically diverse. This is a difficult task as a model trained in one language may perform poorly in others as pointed out by Spiesberger et al \cite{spiesberger23_interspeech}. This 
gives rise to a necessity where individual models may be required for different languages. Maintaining separate models for each language not only increases complexity but also poses a significant challenge in terms of scalability, maintenance, and resource allocation for AAD systems. 
Moreover, it can lead to inconsistencies in performance across languages and hinder the seamless integration of AAD systems into multilingual environments. One way to solve this problem is to train model on the combination of different languages. However, this is not exactly a feasible solution, as training models on the combination of multiple languages will require substantial computational resources and also training data for different languages may not be available at the same time. To mitigate this, we propose, \textbf{CoLLAB} (\textbf{CoLLAB}orative Approach for Multilingual Abuse Detection), a novel framework, that doesn't require training at all and enables seamless integration of models trained in various languages through the process of weight-averaging. This results in a single unified model that demonstrates competitive performance in AAD across multiple languages while comprising the same number of parameters as that of the models trained on individual languages. The proposed framework remains effective even in scenarios where training data for certain languages may not be available at the same time. Notably, \textbf{CoLLAB} offers the added benefit of operating in \textit{Plug-in} mode, facilitating the seamless integration of models trained in various languages alongside those trained on ADIMA. 
This approach ensures consistent performance across upcoming new languages as well as previously seen languages.

\begin{figure}[htbp]
    \centering
    \includegraphics[width=0.45\textwidth, trim=0 0.6cm 0 0, clip]{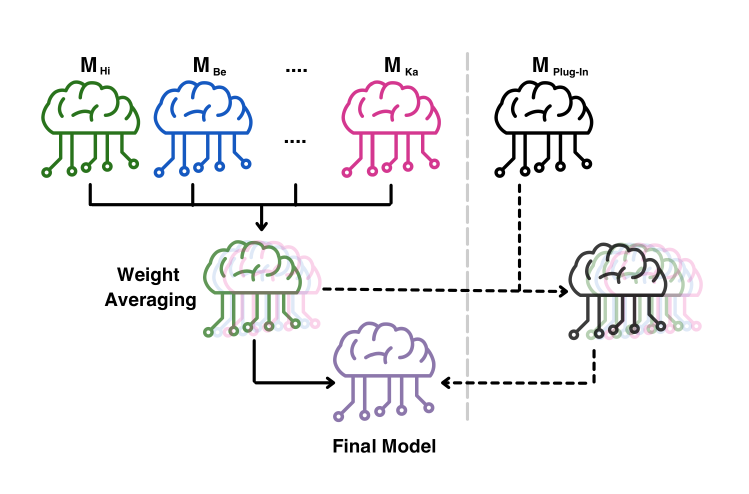}
    \caption{\textbf{CoLLAB}: Here, $M_{Hi}$, $M_{Be}$, $M_{Ka}$ stands for models trained on Hindi, Bengali, Kannada respectively. \textbf{Left:} Models trained on ADIMA languages and weight-averaging of the models' weights for a single unified model; \textbf{Right (Plug-in Mode)} Model trained on external language, that can be integrated into \textbf{CoLLAB} and $M_{Plug-in}$ signifies the Plug-in model}
    \label{fig:collm}
\end{figure}

\noindent \textbf{To summarize, the main contributions are three folds:}
\begin{itemize}
  \item A comprehensive comparative study of different PTM representations to investigate the efficacy of paralingual PTM representations for AAD and our results show that they achieve the topmost performance for different languages.
  \item We show that combining representations from different PTMs leads to improved AAD. See Section \ref{results}.  
  \item We present, \textbf{CoLLAB} (Figure \ref{fig:collm}), a novel unified and collaborative framework that allows merging of models trained in different languages for achieving competitive performance across different languages for AAD. 
\end{itemize}

\noindent For reproducibility of our experiments and as a reference for future studies to build upon \textbf{CoLLAB} for collaborative AAD, we will open source the codes and models built as part of our study after the double-blind review. As \textbf{CoLLAB} in \textit{Plug-in} mode will allow merging of models trained by different researchers in different languages and making it in true sense a collaborative modeling framework for AAD. However, \textbf{CoLLAB} will expect that the researchers follow the same model architectures as given in this study. Kindly refer to Section \ref{collm} for further details. 

\section{Pre-Trained Representations}
In this section, we discuss PTMs whose representations are used in our study. We utilize TRILLsson \cite{shor22_interspeech} as paralingual PTM that was derived from SOTA paralingual conformer CAP12 \cite{shor2022universal} through knowledge distillation. Unlike CAP12, TRILLsson is openly accessible and achieves SOTA performance in the NOSS benchmark. TRILLsson was trained on AudioSet and Libri-light datasets, while CAP12 relies on the YT-U dataset, which potentially may include multilingual data. We employ TRILLsson\footnote{\url{https://tfhub.dev/google/nonsemantic-speech-benchmark/trillsson4/1}} accessible via TensorFlow Hub and returns a vector of 1024-dimension size through aggregrating over time. We employ MMS \cite{pratap2023scaling} and Whisper \cite{radford2023robust} as multilingual PTMs in our study and they have shown SOTA in various downstream multilingual speech processing tasks. MMS was trained in over 1400 languages in a self-supervised manner while Whisper on 96 languages in a weakly-supervised manner. For monolingual PTM, we choose WavLM \cite{chen2022wavlm}, which has achieved SOTA on SUPERB. Additionally, we consider x-vector \cite{8461375}, a SOTA time-delay neural network for speaker recognition. We have choosen x-vector as its representations have shown superior performance for related tasks such as SER \cite{pappagari2020x}, predicting shout intensity \cite{fukumori2023investigating}, depression detection \cite{9746068} where capturing verbal cues such as pitch, intensity, tone, etc. are very important same as in AAD. \par

We experiment with MMS\footnote{\url{https://huggingface.co/facebook/mms-1b}}, Whisper\footnote{\url{https://huggingface.co/openai/whisper-large}}, and WavLM\footnote{\url{https://huggingface.co/microsoft/wavlm-large}} available in Huggingface. We took x-vector\footnote{\url{https://huggingface.co/speechbrain/spkrec-xvect-voxceleb}} from speechbrain \cite{speechbrain} accesible through Huggingface. The input audio samples are sampled to 16kHz before passing through these PTMs and representations of 1280, 1024, 512-dimensional vectors for MMS, WavLM, x-vector are obtained respectively from the last hidden states through pooling average. For Whisper, we extract representations of 1280-dimension size from the encoder through average-pooling and discarding the decoder.

\section{CoLLAB}
\label{collm}
In this section, we present, \textbf{CoLLAB}, a novel framework for abuse detection across multiple languages. The proposed framework is shown in Figure \ref{fig:collm}. \par

\textbf{CoLLAB} operates under the assumption that models trained on different languages share a common architecture. This assumption enables seamless integration and collaboration among models trained in diverse languages within the framework. \textbf{CoLLAB} allows the merging of the trained models through weight-averaging. Let $W_i$ represent the model weights for language $i$, where $i = 1, 2, \ldots, n$ and $n$ is the total number of languages. First, we normalize the weights of the models using L1 normalization given by $\hat{W}_i$. 
\vspace{-0.25cm}
\begin{equation}
\hat{W}_i = \frac{W_i}{\lVert W_i \rVert_1} \label{norm}
\end{equation}
\vspace{-0.28cm}

Normalizing the weights before averaging ensure fair representation, alignment of weight magnitudes, and stability in combining models trained in different languages. Then, we average the normalized model weights and the output is given by $\bar{W}$.
\vspace{-0.3cm}
\begin{equation}
\bar{W} = \sum_{i=1}^{n} \hat{W}_i
\end{equation}
\vspace{-0.28cm}

This formulation ($\bar{W}$) ensures that the resulting unified model captures the collective knowledge from models trained in different languages while keeping the number of parameters the same as the individual models trained in the languages. 

\begin{figure}[htbp]
    \centering
    \includegraphics[width=0.48\textwidth,trim=0 1cm 0 0, clip]{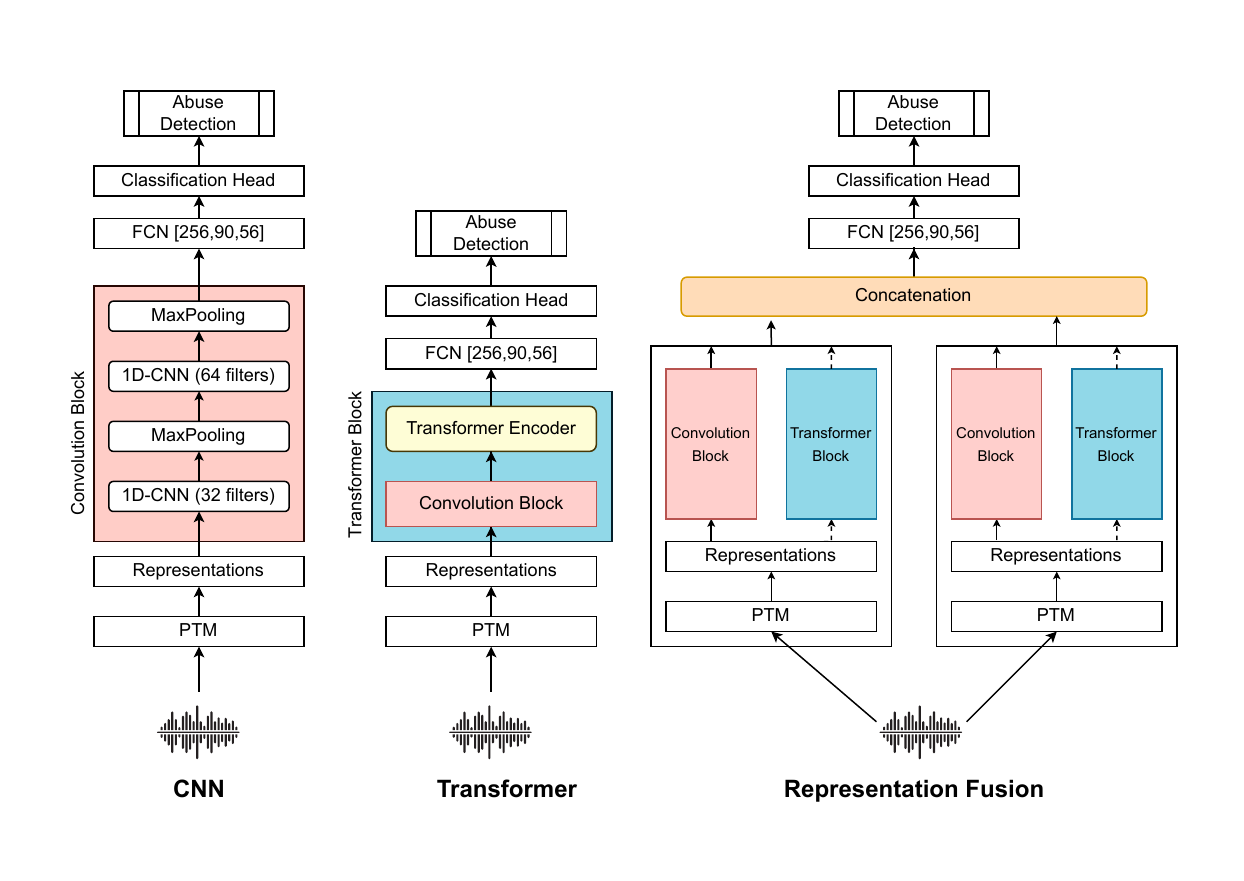}
    \caption{Downstream Model Architectures; Here, FCN stands for Fully Connected Network; Representation Fusion represents the model architecture for combination of PTM representations; We use either convolution or transformer block for the combination of PTM representations without intermixing them.}
    \label{downstream}
\end{figure}

\vspace{-0.3cm}

\section{Experiments}

\begin{table*}[htbp]
\scriptsize
\centering
\caption{Evaluation Scores for models built on different PTM representations; PTM, Down stands for PTM representations, Downstream model; X, Wa, T, M, W stands for x-vector, WavLM, TRILLsson, MMS, Whisper respectively; \textit{Be}, \textit{Bh}, \textit{Gu}, \textit{Ha}, \textit{Hi}, \textit{Ka}, \textit{Ma}, \textit{Od}, \textit{Pu}, \textit{Ta} represents Bengali, Bhojpuri, Gujarati, Haryanvi, Hindi, Kannada, Malayalam, Odiya, Punjabi, Tamil; Acc, F1 stands for Accuracy and F1; \textbf{\textcolor{red}{Red}}, \textbf{\textcolor{blue}{Blue}}, \textbf{\textcolor{violet}{Violet}} represents the colors of the first, second, third scores in descending order for each language and the comparison for a language is done across different models in columnwise manner: The notations used in this Table holds for Table \ref{Tablecomp} and Figure \ref{heatmaps}}
\begin{tabular}{p{0.3cm}p{0.4cm}p{0.27cm}p{0.4cm}p{0.27cm}p{0.4cm}p{0.27cm}p{0.4cm}p{0.27cm}p{0.4cm}p{0.27cm}p{0.4cm}p{0.27cm}p{0.4cm}p{0.27cm}p{0.4cm}p{0.27cm}p{0.4cm}p{0.27cm}p{0.4cm}p{0.27cm}p{0.4cm}}
\toprule
\textbf{PTM} & \textbf{Down} & \multicolumn{2}{c}{\textbf{Be}} & \multicolumn{2}{c}{\textbf{Bh}} & \multicolumn{2}{c}{\textbf{Gu}} & \multicolumn{2}{c}{\textbf{Ha}} & \multicolumn{2}{c}{\textbf{Hi}} & \multicolumn{2}{c}{\textbf{Ka}}& \multicolumn{2}{c}{\textbf{Ma}}& \multicolumn{2}{c}{\textbf{Od}}& \multicolumn{2}{c}{\textbf{Pu}}& \multicolumn{2}{c}{\textbf{Ta}} \\ 
 &  & \text{Acc} & \text{F1} & \text{Acc} & \text{F1} & \text{Acc} & \text{F1} & \text{Acc} & \text{F1} & \text{Acc} & \text{F1} & \text{Acc} & \text{F1}& \text{Acc} & \text{F1}& \text{Acc} & \text{F1}& \text{Acc} & \text{F1}& \text{Acc} & \text{F1} \\ \midrule

\multirow{2}{*}{\textbf{X}} & CNN & 79.45& 78.58 & 76.19 & \textcolor{blue}{\textbf{76.15}} &78.45  &77.51  &80.87  & 80.70 &78.32  & 77.93 &79.94 & 77.87 & 84.41 & 81.23 & 81.37 & 78.33 & 82.83  & 81.54 &81.40 & 80.32\\ 
 & TF & \textcolor{violet}{\textbf{79.73}} & \textcolor{violet}{\textbf{79.61}} & 75.89 & \textcolor{violet}{\textbf{76.01}} & 77.90 & 75.91 & 81.15 & 79.58 & 78.05 & 78.00 & 79.67 & 79.32 & \textcolor{violet}{\textbf{84.68}} & 81.64 & 81.64 & 78.98 &82.56  & 81.75 &80.86&80.01 \\ 
 \midrule

\multirow{2}{*}{\textbf{Wa}} & CNN & 78.38 & 76.62 & 74.40 & 73.81 & 80.11 & 79.24 & 78.96 & 77.19 & 75.33 & 74.89 & 78.32 & 77.68 & 80.11 & 78.65 & 78.08 & 77.03 & 80.93 & 79.48 & 79.24 & 77.89\\ 
 & TF & 79.19 & 76.59 & 74.82 & 73.61 & 77.35 & 76.82 & 77.04 & 75.52 & 75.07 & 74.53 & 80.76 & 79.22 & 79.57 & 78.42 & 78.90 & 77.79 & 81.20 & 80.48 & 78.44 & 78.16\\ 
 \midrule

\multirow{2}{*}{\textbf{T}} & CNN & \textcolor{blue}{\textbf{81.35}} & \textcolor{blue}{\textbf{80.72}} & \textcolor{red}{\textbf{77.10}} & \textcolor{red}{\textbf{76.40}} & \textcolor{red}{\textbf{82.32}} & \textcolor{red}{\textbf{82.28}} &  \textcolor{violet}{\textbf{82.51}} &  \textcolor{violet}{\textbf{82.47}} &  \textcolor{red}{\textbf{83.19}} &  \textcolor{red}{\textbf{82.91}} &  \textcolor{red}{\textbf{83.74}} &  \textcolor{red}{\textbf{83.68}} &  \textcolor{red}{\textbf{86.29}} &  \textcolor{red}{\textbf{86.22}} &  \textcolor{red}{\textbf{83.56}} &  \textcolor{red}{\textbf{83.50}}&  \textcolor{violet}{\textbf{83.38}} &  \textcolor{violet}{\textbf{83.31}} &  \textcolor{red}{\textbf{82.21}}& \textcolor{red}{\textbf{81.93}}\\ 
 & TF & \textcolor{red}{\textbf{82.16}} & \textcolor{red}{\textbf{82.06}} & \textcolor{blue}{\textbf{77.08}} & \textcolor{black}{{75.75}} & \textcolor{blue}{\textbf{81.77}}& \textcolor{violet}{\textbf{81.52}} & \textcolor{red}{\textbf{83.33}} &  \textcolor{blue}{\textbf{83.27}} & 79.40 & 79.38 &  \textcolor{blue}{\textbf{82.11}} &  \textcolor{blue}{\textbf{82.05}} &  \textcolor{blue}{\textbf{84.95}} &  \textcolor{blue}{\textbf{84.89}} &  \textcolor{violet}{\textbf{82.23}} &  \textcolor{violet}{\textbf{82.19}} &  \textcolor{blue}{\textbf{84.47}} &  \textcolor{blue}{\textbf{84.40}} &  \textcolor{violet}{\textbf{81.67}}&80.03\\ 
 \midrule

\multirow{2}{*}{\textbf{M}} & CNN & 78.64 & 77.43 & 74.10 & 74.02 & 80.39 & 80.32 & 78.14 & 78.09 & 76.96 & 76.91 & 79.67 & 79.60 & 81.45 & 81.39 &  \textcolor{blue}{\textbf{82.46}} &  \textcolor{blue}{\textbf{82.40}} & 79.83 & 79.78 & 76.01&75.37\\ 
 & TF & 77.03 & 76.99 & 74.71 & 72.77 & 81.21 & 81.15 & 78.41 & 78.35 & 77.23 & 77.19 & 79.40 & 79.35 & 79.30 & 79.25 & 81.92 & 81.87 & 80.38 & 80.32 & 76.28&76.21\\ 
 \midrule

\multirow{2}{*}{\textbf{W}} & CNN & 79.46 & 79.42 & 75.30 & 74.05 & \textcolor{blue}{\textbf{81.77}} & \textcolor{blue}{\textbf{81.72}} & \textcolor{blue}{\textbf{83.15}} &  \textcolor{violet}{\textbf{82.47}} &  \textcolor{blue}{\textbf{82.65}} &  \textcolor{blue}{\textbf{82.60}} & 81.57 & 81.52 & 82.53 & 82.48 & 81.37 & 81.32 & 82.56 & 82.50 &  \textcolor{blue}{\textbf{81.40}}& \textcolor{blue}{\textbf{81.39}}\\ 
 & TF & 78.38 & 78.34 & \textcolor{violet}{\textbf{76.21}} & 74.65 & \textcolor{violet}{\textbf{81.49}} & 81.16 & \textcolor{red}{\textbf{83.33}} &  \textcolor{red}{\textbf{83.28}} &  \textcolor{violet}{\textbf{80.76}} &  \textcolor{violet}{\textbf{80.73}} &  \textcolor{violet}{\textbf{81.84}} &  \textcolor{violet}{\textbf{81.79}} & 83.60 & \textcolor{violet}{\textbf{83.55}} & 80.55 & 80.51 &  \textcolor{red}{\textbf{85.83}} &  \textcolor{red}{\textbf{85.56}} & 81.13&\textcolor{violet}{\textbf{80.43}}\\ 
 \hline

\end{tabular}
\label{Tablesingle}
\end{table*}

\subsection{Dataset}
We use ADIMA \cite{9746718} for our experiments and it contains 11,775 expert-annotated audio samples in ten Indic languages and spanning over 65 hours. It is evenly distributed across languages and comprises 5,108 abusive and 6,667 non-abusive samples from 6,446 unique Users.

\vspace{-0.3 cm}

\subsection{Downstream Classifier}

We experiment with two downstream networks (CNN, Transformer) extensively used in previous works \cite{rathod23_interspeech, arunkumar22b_interspeech, fang23b_interspeech, campbell23_interspeech} for various speech processing tasks. The model architectures are shown in Figure \ref{downstream}. The extracted representations from the PTMs are passed through the networks. For CNN model, we build a convolution block that contains 1D-CNN layer and maxpooling layer followed by fully connected network (FCN). For Transformer model, we have used the same convolution block as used in CNN model followed by a transformer encoder \cite{vaswani2017attention} with number of the heads as 8. For combination of PTM representations, we pass individual PTM representations through Convolution/Transformer block as shown in Figure \ref{downstream} and then concatenated and passed through FCN for final classification. \par

For the classification head, we use softmax activation function that outputs the probabilities pertaining to each class i.e abuse \textit{vs} non-abuse. We train the models for 50 epochs using Rectified Adam as optimizer with a learning rate of 1e-3 and batch size of 32. We have also used early stopping and dropout for preventing overfitting. We use Accuracy and F1 (macro average) as metrics for evaluating our models. We use Tensorflow library for implementations. We use the official split given by \cite{gupta2022adima} for training and testing our models. For models trained on individual PTM representations, trainable parameters of the models range between 4.2M to 10.5M and with combination of PTM representations it is between 8.4M to 11.3M.

\begin{table*}[htbp]
\scriptsize
\centering
\caption{Evaluation Scores for models built on the combination of different PTM representations; Fusion represents the fusion of different PTM representations; Comparison for a language is done across different models in columnwise manner}
\begin{tabular}
{p{0.45cm}p{0.4cm}p{0.27cm}p{0.4cm}p{0.27cm}p{0.4cm}p{0.27cm}p{0.4cm}p{0.27cm}p{0.4cm}p{0.27cm}p{0.4cm}p{0.27cm}p{0.4cm}p{0.27cm}p{0.4cm}p{0.27cm}p{0.4cm}p{0.27cm}p{0.4cm}p{0.27cm}p{0.4cm}}
\toprule
\textbf{Fusion} & \textbf{Down} & \multicolumn{2}{c}{\textbf{Be}} & \multicolumn{2}{c}{\textbf{Bh}} & \multicolumn{2}{c}{\textbf{Gu}} & \multicolumn{2}{c}{\textbf{Ha}} & \multicolumn{2}{c}{\textbf{Hi}} & \multicolumn{2}{c}{\textbf{Ka}}& \multicolumn{2}{c}{\textbf{Ma}}& \multicolumn{2}{c}{\textbf{Od}}& \multicolumn{2}{c}{\textbf{Pu}}& \multicolumn{2}{c}{\textbf{Ta}} \\ 
 &  & \text{Acc} & \text{F1} & \text{Acc} & \text{F1} & \text{Acc} & \text{F1} & \text{Acc} & \text{F1} & \text{Acc} & \text{F1} & \text{Acc} & \text{F1}& \text{Acc} & \text{F1}& \text{Acc} & \text{F1}& \text{Acc} & \text{F1}& \text{Acc} & \text{F1} \\ \midrule
\multirow{2}{*}{\shortstack{X+T}} & CNN & 81.08 & 80.27 & 77.38 & 74.11 & 80.11 & 79.01 & 82.78 & 82.05 & 80.21 & 80.21 & \textcolor{blue}{\textbf{82.65}} & \textcolor{red}{\textbf{81.84}} & \textcolor{violet}{\textbf{87.36}} & \textcolor{blue}{\textbf{85.75}} & \textcolor{violet}{\textbf{83.56}} & 81.28 & 84.74 & 82.01 & \textcolor{red}{\textbf{87.20}} & \textcolor{blue}{\textbf{83.01}}\\ 
 & TF & 80.54 & 79.72 & 76.79 & 73.12 & \textcolor{violet}{\textbf{82.93}} & \textcolor{red}{\textbf{82.87}} & 80.87 & 80.17 & 80.76 & 80.75 & 80.75 & 77.93 & 85.48 & 81.53 & 83.01 & 82.22 & 85.01 & 85.00 & \textcolor{blue}{\textbf{83.82}} & \textcolor{red}{\textbf{83.28}}\\ \midrule



\multirow{2}{*}{\shortstack{X+W}} & CNN & 78.64 & 78.37 & 77.97 & \textcolor{blue}{\textbf{77.38}} & 82.32 & 79.28 & 82.78 & 81.84 & 78.59 & 77.50 & 80.75 & 76.95 & 85.75 & 81.25 & 82.46 & 81.64 & 83.65 & 82.28 & 82.74 & 82.47\\ 
 & TF & 77.02 & 76.04 & 77.97 & 75.11 & 82.32 & 78.88 & 84.69 & 84.53 & 81.30 & 81.28 & 81.84 & 79.85 & 86.55 & 82.79 & 82.19 & 81.47 & 85.01 & \textcolor{violet}{\textbf{85.01}} & 81.94 & 77.28\\ \midrule


 \multirow{2}{*}{\shortstack{M+T}} & CNN & 80.54 & 79.82 & 77.67 & 74.27 & 81.21 & 79.83 & 82.24 & 80.05 & 81.03 & 80.99 & \textcolor{red}{\textbf{83.74}} & 79.70 & \textcolor{black}{{86.29}} & 82.72 & \textcolor{violet}{\textbf{83.56}} & 81.82 & 85.28 & 84.74 & 82.75 & 82.74 \\ 
 & TF & 79.46  & 78.45 & 76.49 & 72.84 & 78.45 & 74.78 & 81.42 & 81.25 & 78.32 & 78.14 & 81.57 & 78.31 & 83.87 & 78.94 & 81.92 & 81.07 & 83.65 & 83.63 & \textcolor{violet}{\textbf{83.56}} & 79.44 \\ \midrule

 \multirow{2}{*}{\shortstack{M+W}} & CNN & \textcolor{red}{\textbf{82.97}} & \textcolor{violet}{\textbf{81.62}} & 77.67 & \textcolor{violet}{\textbf{76.78}} & \textcolor{blue}{\textbf{83.14}} & 78.86 & {84.97} & 83.87 & \textcolor{violet}{\textbf{82.38}} & \textcolor{violet}{\textbf{81.54}} & 81.03 & \textcolor{violet}{\textbf{80.75}}& 84.40 & \textcolor{violet}{\textbf{83.87}} & \textcolor{red}{\textbf{84.65}} & 82.19 & 84.74 & 84.72 & 80.05 & 79.24\\ 
 & TF & 78.65 & 78.12 & 77.97 & 74.96 & 81.22 & 77.44 & 82.24 & 82.07 & 79.94 & 79.92 & \textcolor{violet}{\textbf{82.38}} & 79.79 & 84.94 & 80.73 & 81.91 & 80.88 & 84.46 & 84.43 & 83.55 & 78.49\\ \midrule

 \multirow{2}{*}{\shortstack{T+Wa}} & CNN & \textcolor{blue}{\textbf{82.70}} & \textcolor{red}{\textbf{82.06}} & \textcolor{violet}{\textbf{78.27}} & 76.19 & \textcolor{black}{\text{82.32}} & 79.00 & 81.96 & 81.42 & 80.21 & 80.20 & \textcolor{blue}{\textbf{82.65}} & 79.55 & 86.55 & \textcolor{blue}{\textbf{85.75}} & 82.74 & 81.64 & 84.74 & 84.46 & {83.28} & 78.50\\ 
 & TF & 81.08 & 80.33 & 75.60 & 72.78 & 80.39 & 76.38 & {84.43} & 84.29 & 80.22 & 80.16 & 81.30 & 78.15 & 85.48 & 81.76 & 80.82 & 79.66 & 85.01 & 84.99 & 81.94 & 77.14\\ \midrule

 \multirow{2}{*}{\shortstack{W+Wa}} & CNN & \textcolor{violet}{\textbf{82.43}} & 80.81 & 77.97 & 75.22 & 81.49 & \textcolor{blue}{\textbf{80.66}} & \textcolor{violet}{\textbf{85.24}} & 85.10 & \textcolor{red}{\textbf{83.73}} & \textcolor{red}{\textbf{83.68}} & \textcolor{violet}{\textbf{82.38}} & 80.30 & 86.55 & 82.10& \textcolor{blue}{\textbf{84.11}} & \textcolor{violet}{\textbf{82.73}} & 84.19 & 83.37 & 83.01 & 81.40\\ 
 & TF & 78.92 & 78.08 & \textcolor{blue}{\textbf{78.87}} & 76.17 & 82.04 & 78.26 & \textcolor{blue}{\textbf{85.25}} & \textcolor{violet}{\textbf{85.11}} & 80.22 & 80.21 & 81.30 & 79.47 & 84.68 & 80.44 & 82.74 & 82.04 & \textcolor{violet}{\textbf{85.29}} & \textcolor{blue}{\textbf{85.28}} & 80.32 & 75.82\\ \midrule

 \multirow{2}{*}{\shortstack{W+T}} & CNN & \textcolor{blue}{\textbf{82.70}} & \textcolor{blue}{\textbf{81.72}} & \textcolor{red}{\textbf{80.35}} & \textcolor{red}{\textbf{78.27}} & \textcolor{red}{\textbf{83.42}} & \textcolor{violet}{\textbf{80.38}} & \textcolor{red}{\textbf{85.79}} & \textcolor{red}{\textbf{85.60}} & 82.11 & 80.75 & 81.84 & \textcolor{blue}{\textbf{81.02}} & \textcolor{red}{\textbf{87.90}} & 82.27 & \textcolor{blue}{\textbf{84.11}} & \textcolor{red}{\textbf{83.83}} & \textcolor{red}{\textbf{86.37}} & 84.19 & {83.28} & \textcolor{violet}{\textbf{82.74}} \\ 
 & TF & 81.08 & 80.20 & 77.98 & 74.23 & 79.83 & 76.10 & \textcolor{blue}{\textbf{85.25}} & \textcolor{blue}{\textbf{85.14}} & \textcolor{blue}{\textbf{82.93}} & \textcolor{blue}{\textbf{82.11}} & 82.11 & 78.84 & \textcolor{blue}{\textbf{87.63}}& \textcolor{red}{\textbf{87.36}} & \textcolor{blue}{\textbf{84.11}} & \textcolor{blue}{\textbf{83.28}} & \textcolor{blue}{\textbf{86.33}} & \textcolor{red}{\textbf{85.56}} & 82.21 & 77.55 \\ \hline

\end{tabular}
\label{Tablecomp}
\end{table*}

\begin{table*}[htbp]
\scriptsize
\centering
\caption{Evaluation Scores with \textbf{CoLLAB}; For \textbf{CoLLAB}, we experiment with the models showing top performance in  Table \ref{Tablesingle} (CNN with TRILLsson) and Table \ref{Tablecomp} (CNN with Whisper + TRILLsson); 
D represents the difference between the original scores on a language and its corresponding score after \textbf{CoLLAB}; Scores are accuracy scores}
\begin{tabular}{p{0.45cm}p{0.4cm}p{0.3cm}p{0.38cm}p{0.3cm}p{0.38cm}p{0.3cm}p{0.38cm}p{0.3cm}p{0.38cm}p{0.3cm}p{0.38cm}p{0.3cm}p{0.38cm}p{0.3cm}p{0.38cm}p{0.3cm}p{0.38cm}p{0.3cm}p{0.38cm}p{0.3cm}p{0.4cm}}
\toprule
\textbf{PTM} & \textbf{Down} & \multicolumn{2}{c}{\textbf{Be}} & \multicolumn{2}{c}{\textbf{Bh}} & \multicolumn{2}{c}{\textbf{Gu}} & \multicolumn{2}{c}{\textbf{Ha}} & \multicolumn{2}{c}{\textbf{Hi}} & \multicolumn{2}{c}{\textbf{Ka}}& \multicolumn{2}{c}{\textbf{Ma}}& \multicolumn{2}{c}{\textbf{Od}}& \multicolumn{2}{c}{\textbf{Pu}}& \multicolumn{2}{c}{\textbf{Ta}} \\ 
 &  & \text{\textbf{C}} & \text{D} & \text{C} & \text{D} & \text{C} & \text{D} & \text{C} & \text{D} & \text{C} & \text{D} & \text{C} & \text{D}& \text{C} & \text{D}& \text{C} & \text{D}& \text{C} & \text{D}& \text{C} & \text{D} \\ \midrule
\multirow{1}{*}{\shortstack{T}} & CNN & 79.20 &2.15& 76.12 &0.98& 80.13 &2.19& 81.23 &1.28& 81.41 &1.78& 82.48 &1.26& 83.37 &2.92& 82.03 &1.53& 81.10 &2.28& 79.83&2.38\\
 \midrule

\multirow{1}{*}{\shortstack{W+T}} & CNN & 80.48 & 2.22 & 78.69 & 1.66 & 80.45 & 2.97 & 83.13  & 2.66 & 81.19 & 0.92 & 81.80 & 0.04 & 84.01 & 3.89 & 82.91 & 1.20 & 82.47 & 3.90 & 80.05 & 3.23\\ 
\midrule
\end{tabular}
\label{modelcollm}
\end{table*}
\vspace{-0.3cm}
\subsection{Experimental Results}

\label{results}
\noindent\textbf{Within-language:} First, we discuss the results obtained by training models with individual PTM representations shown in Table \ref{Tablesingle}. The TRILLsson (Paralingual PTM) representations demonstrated the best performance across most languages, showcasing their effectiveness in capturing verbal cues crucial for AAD. Whisper representations followed closely, showing the second-best performance in most languages and leading in couples, likely due to exposure to diverse linguistic contexts during its pre-training. 
x-vector, WavLM, and MMS shows mixed performance. Amongst downstream networks, CNN shows better performance than transformer. 
Results with TRILLsson improve over previous works leveraging single PTM representations \cite{gupta2022adima} and also \cite{spiesberger23_interspeech}, that uses handcrafted features for AAD. However, due to space constraints, we are not able to present the comparison table. \par

Table \ref{Tablecomp} shows the results when PTM representations are combined. We notice improvements in most instances compared to built on individual PTM representations. We experimented with the combinations where either TRILLsson and Whisper is present as these two PTMs showed the best performance individually. Combining TRILLsson and Whisper performed the topmost among all the other combinations showing the best complementary behavior. 

\begin{figure}[t!]
    \centering
    \subfloat[T]{\includegraphics[width=0.49\linewidth]{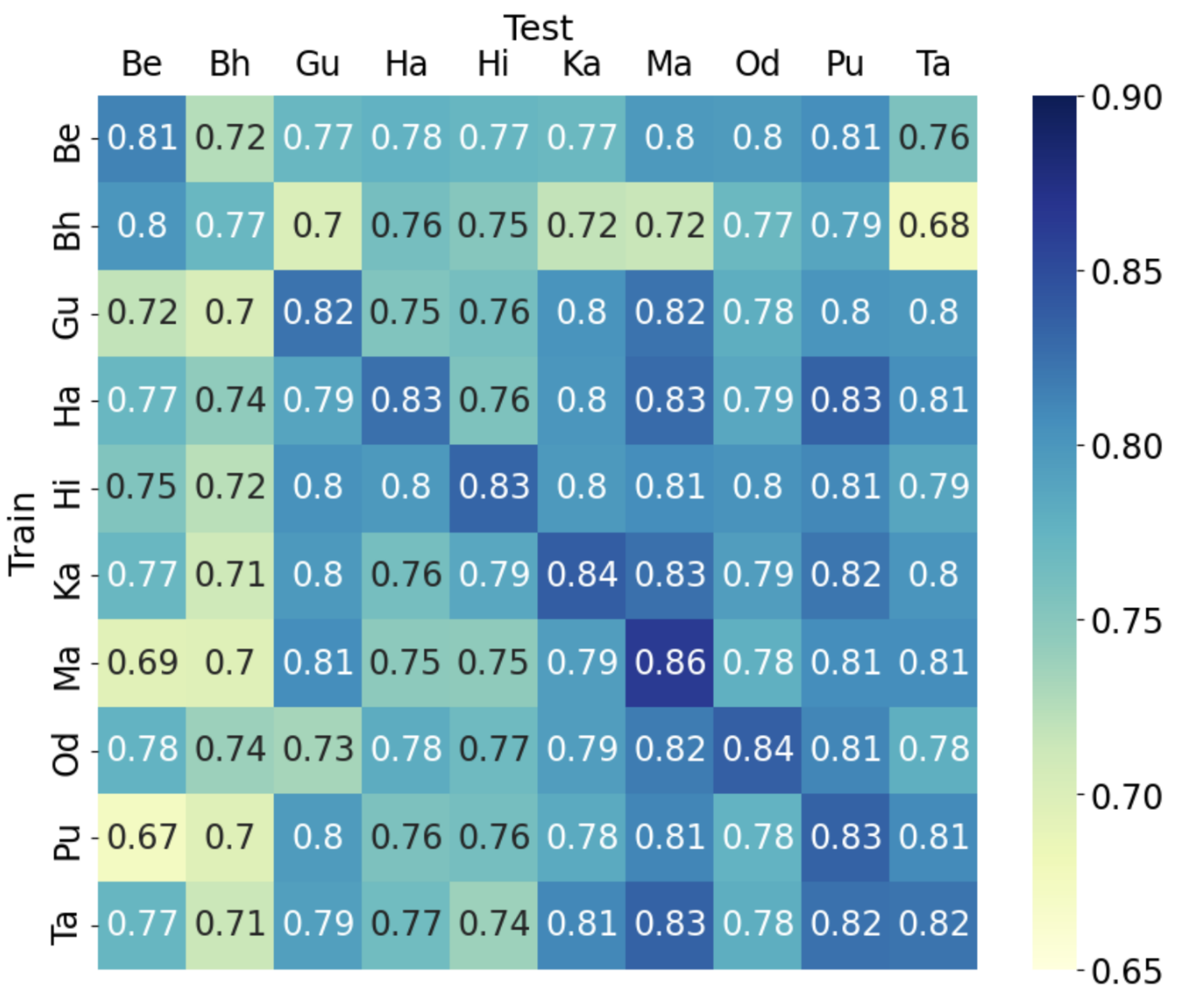}}\label{ttt}
    \subfloat[W+T]{\includegraphics[width=0.49\linewidth]{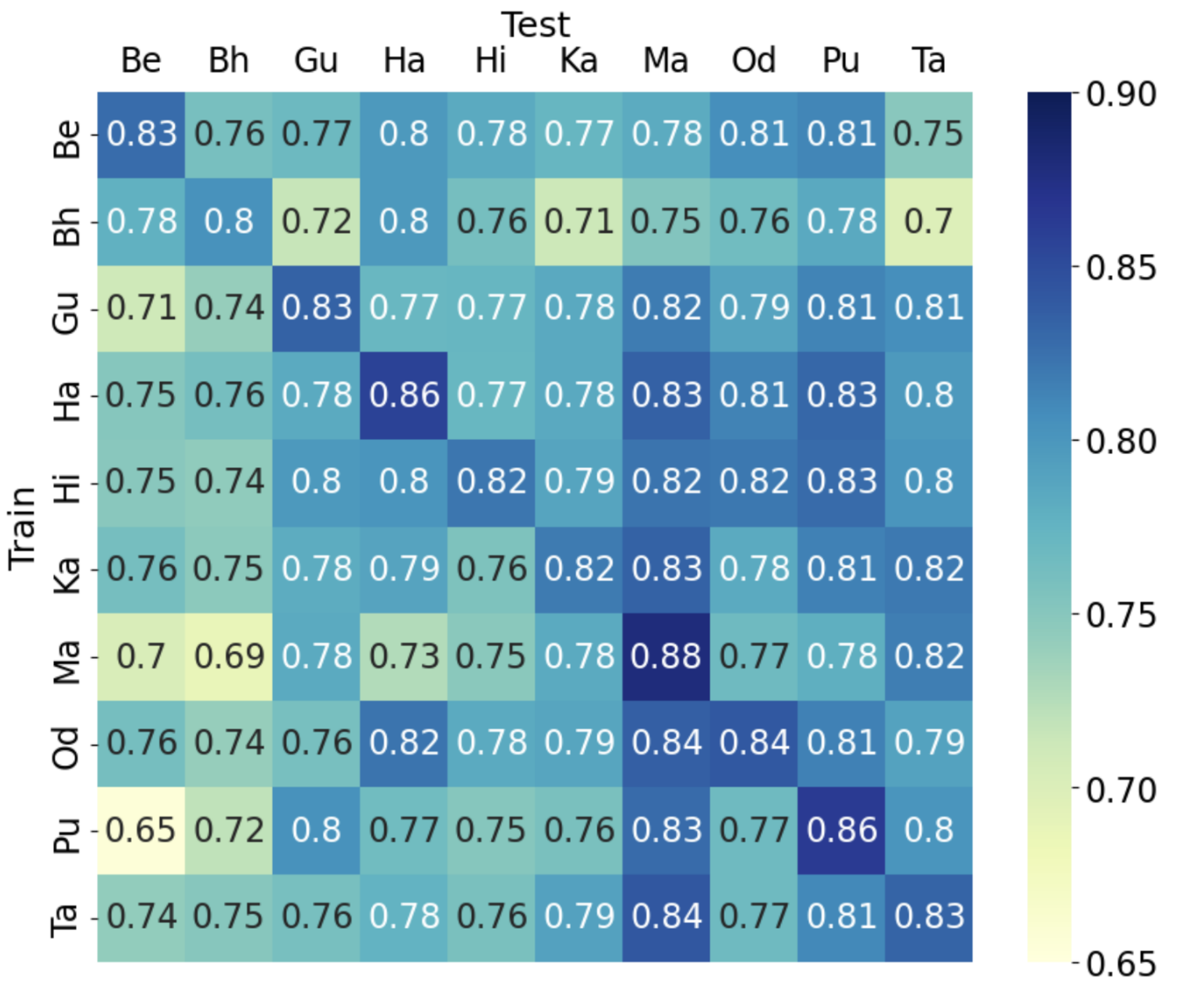}}\label{fig:4}
    \caption{Cross-lingual Evaluation Scores; Train, Test represents the language trained on and the languages tested on; Here, the values are accuracy scores and rounded off to two decimal points}
    \label{heatmaps}
\end{figure}

\noindent\textbf{Cross-lingual:} The results are shown in Figure \ref{heatmaps}. We experiment with only the best models from Table \ref{Tablesingle} (CNN with TRILLsson) and Table \ref{Tablecomp} (CNN with Whisper + TRILLsson). Despite these models achieving top performance when trained and evaluated in the same language, their cross-lingual performance is low. For example, models trained in Punjabi (Figure \ref{heatmaps}) perform poorly in most other languages and also models trained in other languages performs poorly in Punjabi. This low cross-lingual generalizability issue persists with both models trained with individual and combined PTM representations and calls for individual models to be trained in specific languages.  

\noindent\textbf{With CoLLAB:} The scores obtained by the single unified model as a resultant of \textbf{CoLLAB} are presented in Table \ref{modelcollm}. We observe consistent performance across all the languages. This eliminates the need for training individual models for each language, thereby resolving scalability, maintenance, and resource allocation concerns. We also notice slight decrease in performance compared to models trained and evaluated in the same language (shown in D column in Figure \ref{modelcollm}) and we aim to look into this in future work. However, this presents a tradeoff, we have to incur for a single unified model that performs competitively across different languages as in real-world settings, we would want a model show relatively good performance across multiple languages and not in a individual language and performs poorly in other languages.

\section{Conclusion}
In this, we show that representations from TRILLsson (paralingual PTM) attain the topmost performance for AAD in comparison to other PTM (Whisper, MMS, WavLM, x-vector) representations. This performance is attributed to their effectiveness in capturing verbal cues such as pitch, tone, and intensity which are important attributes for AAD much more effectively than other representations. Additionally, we show that combining PTM representations leads to further improvement in AAD performance. Despite this, challenges persist in achieving cross-lingual generalizability, with certain languages requiring language-specific training giving rise to scalability, maintenance, and resource allocation issues and hindering the practical deployment of AAD systems in linguistically diverse real-world scenarios. To address this, we propose \textbf{CoLLAB}, a novel framework that facilitates the merging of models trained in different languages and results in a single unified model exhibiting competitive AAD performance across different languages. Our work can guide future studies, firstly, in selecting the most suitable representations for AAD and secondly, act as a reference for the development of collaborative unified frameworks not only for AAD but also for related applications for deployment in linguistically diverse real-world environments. 

\bibliographystyle{IEEEtran}
\bibliography{main.bib}

\begin{thebibliography}{10}
\providecommand{\url}[1]{#1}
\csname url@samestyle\endcsname
\providecommand{\newblock}{\relax}
\providecommand{\bibinfo}[2]{#2}
\providecommand{\BIBentrySTDinterwordspacing}{\spaceskip=0pt\relax}
\providecommand{\BIBentryALTinterwordstretchfactor}{4}
\providecommand{\BIBentryALTinterwordspacing}{\spaceskip=\fontdimen2\font plus
\BIBentryALTinterwordstretchfactor\fontdimen3\font minus \fontdimen4\font\relax}
\providecommand{\BIBforeignlanguage}[2]{{%
\expandafter\ifx\csname l@#1\endcsname\relax
\typeout{** WARNING: IEEEtran.bst: No hyphenation pattern has been}%
\typeout{** loaded for the language `#1'. Using the pattern for}%
\typeout{** the default language instead.}%
\else
\language=\csname l@#1\endcsname
\fi
#2}}
\providecommand{\BIBdecl}{\relax}
\BIBdecl

\bibitem{davidson2017automated}
T.~Davidson, D.~Warmsley, M.~Macy, and I.~Weber, ``Automated hate speech detection and the problem of offensive language,'' in \emph{Proceedings of the international AAAI conference on web and social media}, vol.~11, no.~1, 2017, pp. 512--515.

\bibitem{madhu2023detecting}
H.~Madhu, S.~Satapara, S.~Modha, T.~Mandl, and P.~Majumder, ``Detecting offensive speech in conversational code-mixed dialogue on social media: A contextual dataset and benchmark experiments,'' \emph{Expert Systems with Applications}, vol. 215, p. 119342, 2023.

\bibitem{alcantara2020offensive}
C.~Alc{\^a}ntara, V.~Moreira, and D.~Feijo, ``Offensive video detection: dataset and baseline results,'' in \emph{Proceedings of the Twelfth Language Resources and Evaluation Conference}, 2020, pp. 4309--4319.

\bibitem{gao2020offensive}
Z.~Gao, S.~Yada, S.~Wakamiya, and E.~Aramaki, ``Offensive language detection on video live streaming chat,'' in \emph{Proceedings of the 28th International Conference on Computational Linguistics}, 2020, pp. 1936--1940.

\bibitem{9746718}
V.~Gupta, R.~Sharon, R.~Sawhney, and D.~Mukherjee, ``Adima: Abuse detection in multilingual audio,'' in \emph{ICASSP 2022 - 2022 IEEE International Conference on Acoustics, Speech and Signal Processing (ICASSP)}, 2022, pp. 6172--6176.

\bibitem{sharon22_interspeech}
R.~Sharon, H.~Shah, D.~Mukherjee, and V.~Gupta, ``{Multilingual and Multimodal Abuse Detection},'' in \emph{Proc. Interspeech 2022}, 2022, pp. 4631--4635.

\bibitem{thakran23_interspeech}
Y.~Thakran and V.~Abrol, ``{Investigating Acoustic Cues for Multilingual Abuse Detection},'' in \emph{Proc. INTERSPEECH 2023}, 2023, pp. 3642--3646.

\bibitem{spiesberger23_interspeech}
A.~A. Spiesberger, A.~Triantafyllopoulos, I.~Tsangko, and B.~W. Schuller, ``{Abusive Speech Detection in Indic Languages Using Acoustic Features},'' in \emph{Proc. INTERSPEECH 2023}, 2023, pp. 2683--2687.

\bibitem{shor2022universal}
J.~Shor, A.~Jansen, W.~Han, D.~Park, and Y.~Zhang, ``Universal paralinguistic speech representations using self-supervised conformers,'' in \emph{ICASSP 2022-2022 IEEE International Conference on Acoustics, Speech and Signal Processing (ICASSP)}.\hskip 1em plus 0.5em minus 0.4em\relax IEEE, 2022, pp. 3169--3173.

\bibitem{arunkumar22b_interspeech}
A.~Arunkumar, V.~{Nileshkumar Sukhadia}, and S.~Umesh, ``{Investigation of Ensemble features of Self-Supervised Pretrained Models for Automatic Speech Recognition},'' in \emph{Proc. Interspeech 2022}, 2022, pp. 5145--5149.

\bibitem{shor22_interspeech}
J.~Shor and S.~Venugopalan, ``{TRILLsson: Distilled Universal Paralinguistic Speech Representations},'' in \emph{Proc. Interspeech 2022}, 2022, pp. 356--360.

\bibitem{pratap2023scaling}
V.~Pratap, A.~Tjandra, B.~Shi, P.~Tomasello, A.~Babu, S.~Kundu, A.~Elkahky, Z.~Ni, A.~Vyas, M.~Fazel-Zarandi \emph{et~al.}, ``Scaling speech technology to 1,000+ languages,'' \emph{arXiv preprint arXiv:2305.13516}, 2023.

\bibitem{radford2023robust}
A.~Radford, J.~W. Kim, T.~Xu, G.~Brockman, C.~McLeavey, and I.~Sutskever, ``Robust speech recognition via large-scale weak supervision,'' in \emph{International Conference on Machine Learning}.\hskip 1em plus 0.5em minus 0.4em\relax PMLR, 2023, pp. 28\,492--28\,518.

\bibitem{chen2022wavlm}
S.~Chen, C.~Wang, Z.~Chen, Y.~Wu, S.~Liu, Z.~Chen, J.~Li, N.~Kanda, T.~Yoshioka, X.~Xiao \emph{et~al.}, ``Wavlm: Large-scale self-supervised pre-training for full stack speech processing,'' \emph{IEEE Journal of Selected Topics in Signal Processing}, vol.~16, no.~6, pp. 1505--1518, 2022.

\bibitem{8461375}
D.~Snyder, D.~Garcia-Romero, G.~Sell, D.~Povey, and S.~Khudanpur, ``X-vectors: Robust dnn embeddings for speaker recognition,'' in \emph{2018 IEEE International Conference on Acoustics, Speech and Signal Processing (ICASSP)}, 2018, pp. 5329--5333.

\bibitem{pappagari2020x}
R.~Pappagari, T.~Wang, J.~Villalba, N.~Chen, and N.~Dehak, ``x-vectors meet emotions: A study on dependencies between emotion and speaker recognition,'' in \emph{ICASSP 2020-2020 IEEE International Conference on Acoustics, Speech and Signal Processing (ICASSP)}.\hskip 1em plus 0.5em minus 0.4em\relax IEEE, 2020, pp. 7169--7173.

\bibitem{fukumori2023investigating}
T.~Fukumori, T.~Ishida, and Y.~Yamashita, ``Investigating the effectiveness of speaker embeddings for shout intensity prediction,'' in \emph{2023 Asia Pacific Signal and Information Processing Association Annual Summit and Conference (APSIPA ASC)}.\hskip 1em plus 0.5em minus 0.4em\relax IEEE, 2023, pp. 1838--1842.

\bibitem{9746068}
J.~V. Egas-López, G.~Kiss, D.~Sztahó, and G.~Gosztolya, ``Automatic assessment of the degree of clinical depression from speech using x-vectors,'' in \emph{ICASSP 2022 - 2022 IEEE International Conference on Acoustics, Speech and Signal Processing (ICASSP)}, 2022, pp. 8502--8506.

\bibitem{speechbrain}
M.~Ravanelli, T.~Parcollet, P.~Plantinga, A.~Rouhe, S.~Cornell, L.~Lugosch, C.~Subakan, N.~Dawalatabad, A.~Heba, J.~Zhong, J.-C. Chou, S.-L. Yeh, S.-W. Fu, C.-F. Liao, E.~Rastorgueva, F.~Grondin, W.~Aris, H.~Na, Y.~Gao, R.~D. Mori, and Y.~Bengio, ``{SpeechBrain}: A general-purpose speech toolkit,'' 2021, arXiv:2106.04624.

\bibitem{rathod23_interspeech}
S.~Rathod, M.~Charola, A.~Vora, Y.~Jogi, and H.~A. Patil, ``{Whisper Features for Dysarthric Severity-Level Classification},'' in \emph{Proc. INTERSPEECH 2023}, 2023, pp. 1523--1527.

\bibitem{fang23b_interspeech}
Y.~Fang, X.~Xing, X.~Xu, and W.~Zhang, ``{Exploring Downstream Transfer of Self-Supervised Features for Speech Emotion Recognition},'' in \emph{Proc. INTERSPEECH 2023}, 2023, pp. 3627--3631.

\bibitem{campbell23_interspeech}
E.~L. Campbell, J.~Dineley, P.~Conde, F.~Matcham, K.~M. White, C.~Oetzmann, S.~Simblett, S.~Bruce, A.~A. Folarin, T.~Wykes, S.~Vairavan, R.~J.~B. Dobson, L.~Docio-Fernandez, C.~Garcia-Mateo, V.~A. Narayan, M.~Hotopf, and N.~Cummins, ``{Classifying depression symptom severity: Assessment of speech representations in personalized and generalized machine learning models.}'' in \emph{Proc. INTERSPEECH 2023}, 2023, pp. 1738--1742.

\bibitem{vaswani2017attention}
A.~Vaswani, N.~Shazeer, N.~Parmar, J.~Uszkoreit, L.~Jones, A.~N. Gomez, {\L}.~Kaiser, and I.~Polosukhin, ``Attention is all you need,'' \emph{Advances in neural information processing systems}, vol.~30, 2017.

\bibitem{gupta2022adima}
V.~Gupta, R.~Sharon, R.~Sawhney, and D.~Mukherjee, ``Adima: Abuse detection in multilingual audio,'' in \emph{ICASSP 2022-2022 IEEE International Conference on Acoustics, Speech and Signal Processing (ICASSP)}.\hskip 1em plus 0.5em minus 0.4em\relax IEEE, 2022, pp. 6172--6176.

\end{thebibliography}

\end{document}